\documentclass[traditabstract]{aa} % for the abstract without structuration
\pdfoutput=1
\newcommand{\ergps}{erg\thinspace s$^{-1}$}
\newcommand{\ergpspsqcm}{erg\thinspace s$^{-1}$\thinspace cm$^{-2}$}
\newcommand{\psqcm}{cm$^{-2}$}
\newcommand{\nH}{$N_{\rm H}$}

\newcommand{\phpspsqcm}{ph\thinspace s$^{-1}$\thinspace cm$^{-2}$}
\newcommand{\cps}{ct\thinspace s$^{-1}$}

\usepackage{graphicx}
\usepackage{txfonts}
\begin{document}
   \title{The Suzaku broadband X-ray spectrum of
the dwarf Seyfert galaxy NGC4395}

%   \subtitle{I. Overviewing the $\kappa$-mechanism}

   \author{K. Iwasawa
          \inst{1}\thanks{Present address: ICREA Research Professor at Institut del Ci\`encies del Cosmos, Universitat de Barcelona, Mart\'i i Franqu\`es, 1, 08028 Barcelona, Spain,
email: {\tt kazushi.iwasawa@icc.ub.edu}
}
\and
          Y. Tanaka
          \inst{2}
          \and
          L.C. Gallo
          \inst{3}
          }
   \institute{INAF-Ossservatorio Astronomico di Bologna, Via Ranzani, 1,
     40127 Bologna, Italy\\
              \email{kazushi.iwasawa@oabo.inaf.it}
\and
              Max-Planck-Institut f\"ur extraterrestrische Physik,
              Giessenbachstra\ss e 1, 85748 Garching, Germany\\
              \email{ytanaka@mpe.mpg.de}
         \and
         Department of Astronomy and Physics, Saint Mary's University,
         Halifax, NS B3H 3C3, Canada\\
             \email{lgallo@ap.smu.ca}
             }

%   \date{Received September 15, 1996; accepted March 16, 1997}

%\abstract{}{}{}{}{}
% 5 {} token are mandatory

\abstract{ 

  We present a Suzaku observation of the dwarf Seyfert galaxy NGC4395
  with an estimated black hole mass of $\sim 10^5 M_{\odot}$. Rapid
  and strong X-ray variability with an rms amplitude of $\sim 60$ per
  cent is observed in the 0.4-10 keV band with the XIS cameras. The
  shape of the light curve appears to depend on energies. The hard
  X-ray emission is detected up to 35 keV with the HXD-PIN detector at
  a similar flux level as observed with the INTEGRAL IBIS. The X-ray
  spectrum below 10 keV is strongly absorbed by partially ionized
  ($\xi\sim 35$ erg\thinspace s\thinspace cm$^{-1}$) gas with a mean
  equivalent hydrogen column density of $\sim 2\times 10^{22}$\psqcm,
  when a simple absorption model is applied. The spectral shape is
  also strongly variable but not a simple function of the source
  brightness. The spectral variability appears to be accounted for
  mainly by continuum slope changes, but variability in the ionized
  absorber may also play some part. The apparently flat spectral slope
  of $\Gamma\simeq 1.4$ below 10 keV, obtained after correcting for
  absorption, is marginally inconsistent with the $\Gamma\sim 2$
  inferred from the 14-35 keV PIN spectrum. If the true spectral slope
  had been as steep as that measured in the hard X-ray band, there
  would have been an extra absorption component, which we are unable
  to detect. Combined with the INTEGRAL measurements, the hard X-ray
  emission above 10 keV exceeds the optical emission in terms of
  luminosity and dominates the broadband energy output, unless a large
  excess of UV disk emission is yet to be detected in the unobservable
  band. A weak Fe K line is seen at 6.4 keV with the average
  equivalent width of 110 eV, which does not show significant flux
  changes over the 3-day observation.}  
   \maketitle
%
%________________________________________________________________

\section{Introduction}

NGC4395 is a dwarf galaxy located at the distance of 4.0 Mpc (Thim et
al 2004) that is known to host a very low luminosity Seyfert nucleus
(Filippenko, Ho \& Sargent 1993). The active nucleus is located at the
centre of the galaxy, which has no obvious bulge component. Unlike
LINER-like, low-luminosity AGN, the optical/UV spectrum of NGC4395
shows broad Balmer emission and high-excitation lines, reminiscent of
genuine Seyfert nuclei but of higher luminosity (e.g., Ho, Filippenko
\& Sargent 1997). Using the relation usually applied to higher
luminosity AGN (Kaspi et al 2000), the bolometric luminosity estimated
from the 5100 \AA\ luminosity is $5.4\times 10^{40}$\ergps (Peterson
et al 2005), which can be powered by a black hole of a few 100
$M_{\odot}$ if it accretes close to the Eddington limit.

\begin{figure*}
\centerline{\includegraphics[width=0.8\textwidth,angle=0]{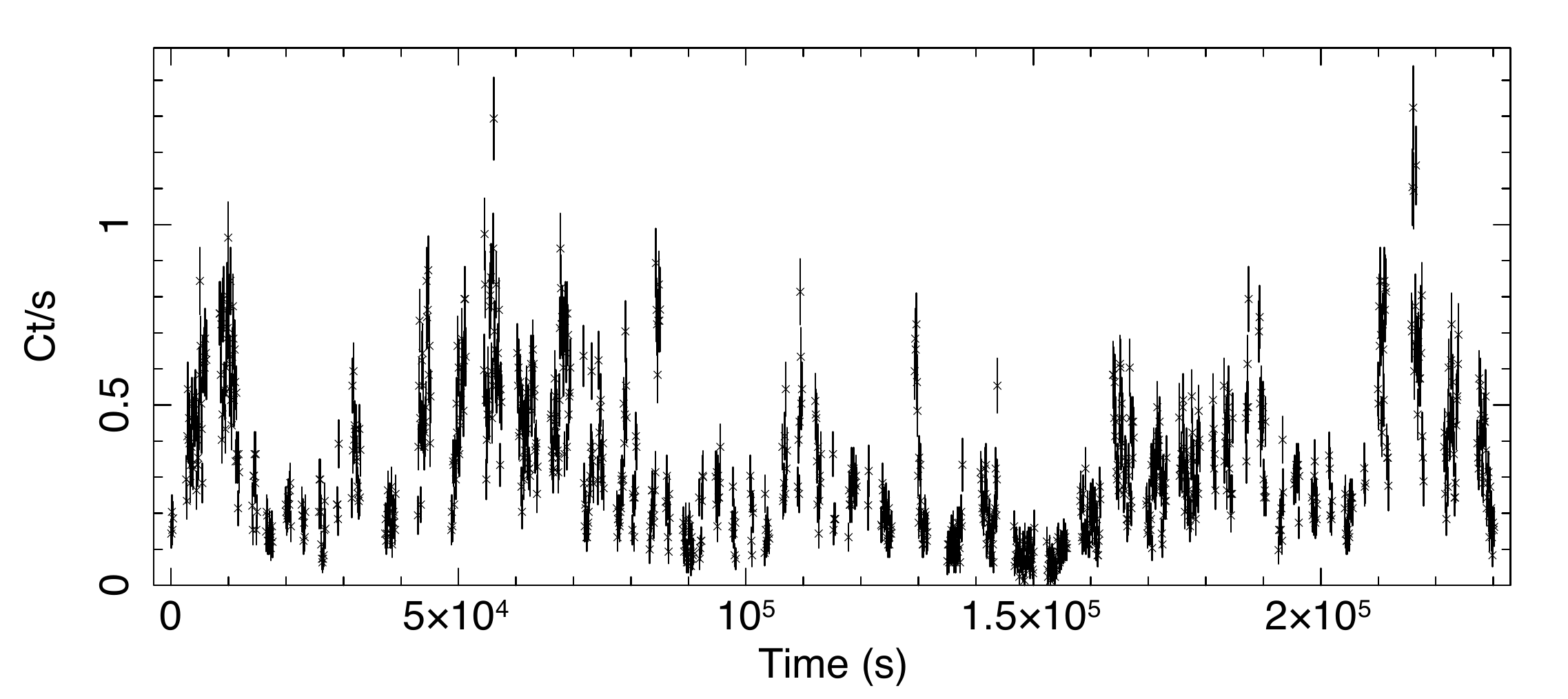}}
%\centerline{\includegraphics[width=0.8\textwidth,angle=0]{12431fg1.ps}}
\caption{ The X-ray light curve of NGC4395 during the Suzaku
  observation. The background-subtracted counts in the 0.4-10 keV band
  from the three XISs are summed together. Each timebin is of a
  100 s duration.}
\end{figure*}

There have been indications that the mass of the black hole in NGC4395
may be $10^5 M_{\odot}$ or lower, including by X-ray timing analysis
(Kraemer et al 1999; Lira et al 1999). The mass derived from the
reverberation mapping technique is found to be $(3.6\pm 1.1)\times
10^5 M_{\odot}$ (Peterson et al 2005). However, the low bulge velocity
dispersion $\leq 30$ km s$^{-1}$ (Filippenko \& Ho 2003) deviates from
the extrapolation of the $M$-$\sigma$ relation (e.g., Tremaine et al
2002) when this estimate of the black hole mass is adopted. Some
researchers (e.g., McHardy et al 2006) propose a lower mass, of even an
order of magnitude below the value obtained from reverberation
mapping.

As pointed out by Peterson et al (2005), the Eddington ratio ($L_{\rm
  bol}/L_{\rm Edd}$) is low $1.3\times 10^{-3}[M_{\rm BH}/(3.6\times
10^5 M_{\odot})]^{-1}$. Provided that the estimate of the bolometric
luminosity derived from the 5100 \AA\ luminosity is correct, even for
the lower mass estimate ($\sim 10^4 M_{\odot}$), the accretion rate
is the order of $10^{-2}$ of the Eddington limit. In terms of the
accretion rate, the X-ray source in NGC4395 corresponds to the
low/hard state of X-ray binaries.

The soft X-ray emission observed with ROSAT (Moran et al 1999,
Lira et al 1999) is weak largely because of strong absorption (Iwasawa et
al 2000). The absorbing material is partially ionized (warm
absorber) and neutral (cold) absorption is usually small (a few times
$10^{20}$ \psqcm\ in \nH). However, an occasional increase in the cold
absorbing column, perhaps due to a small cloud passing across the line
of sight, has been observed in one of the XMM-Newton observations
(Dewangan et al 2009).

The X-ray emission is known to be
extremely variable on short timescales. The temporal properties of
the X-ray emission of NGC4395 was studied in detail using a long
XMM-Newton observation with a consecutive 90 ks duration (Vaughan et
al 2005). The power spectrum of the flux variation shows a break at
$\sim 2\times 10^{-3}$ Hz, which is much higher than those for other
AGN, suggesting a low black hole mass. The rms amplitude of the soft
X-ray flux variability exceeds 100 per cent.

In comparison with other AGN, the X-ray spectrum of NGC4395 below 10
keV appears to be very hard even after correcting for absorption,
$\Gamma\sim 1.4$ (e.g., Shih et al 2004; Moran et al (2005) suggested
an even lower value of $\Gamma \sim 0.6$). Despite the relatively low
2-10 keV flux ($\sim 4\times 10^{-13}$ \ergpspsqcm), NGC4395 was
detected by INTEGRAL IBIS (Beckman et al 2006; Bird et al 2007) in
20-40 keV and 40-100 keV bands. This high-energy X-ray detection may
imply that the hard spectrum continues into the hard band. The
measurement of the spectral shape above 10 keV, where it is less
affected by absorption, is important to assess the origin of the hard
intrinsic spectrum. We observed NGC4395 with Suzaku to measure the
spectral shape above 10 keV and characterise the flux and spectral
variability in the softer band by using the X-ray CCD spectrometer over a
continuous period of more than $\sim 2.5$ days.

\section{Observation and data reduction}

%In the 0.4-1 keV, the total counts from S1 and S0+S3 are 1.4:1.

NGC4395 was observed with Suzaku (Mitsuda et al 2006) on 2007 June 02
-- 05, for a total duration of 230 ks. The XIS cameras were
operated with the Space-Row Charge Injection in the standard observation
mode. The data were taken by the three XISs: two FI CCD cameras (XIS0
and XIS3), one BI CCD camera (XIS1), and the HXD-PIN. The useful
exposures obtained from the XIS and the HXD-PIN were 101 ks and 95
ks, respectively. NGC4395 was observed in the ``HXD nominal'' position
on the detector plane, which is optimised for the maximum throughput
to the HXD rather than the XIS.

The data from version 2
processing\footnote{http://www.astro.isas.ac.jp/suzaku/process/caveats}
were used in the analysis. The cleaned event files produced by the
data processing with the standard data selection criteria were
used. The non X-ray background of the PIN detector was estimated using
the ``tuned'' background
model\footnote{http://www.astro.isas.ac.jp/suzaku/analysis/hxd/pinnxb/tuned/}.
The background-corrected count rates observed with the three XIS in
the 0.4-10 keV band are $0.107\pm 0.001$ (XIS0), $0.116\pm 0.001$
(XIS1), $0.109\pm 0.001$ (XIS3), and $0.031\pm 0.002$ with the HXD-PIN
in the 14-40 keV band.

\section{Results}

\subsection{X-ray variability in the XIS band}

The X-ray source in NGC4395 is strongly variable during the Suzaku
observation, as previously observed with other instruments. Since there are many irregular
interruptions in data sampling because of the low-Earth satellite orbit, the
Suzaku data are not ideal for a detailed timing study 
of this rapidly variable object on short timescales. We refer to Vaughan et al
(2005) for such a study with the uninterrupted 90-ks XMM-Newton
data. Despite the low observation
efficiency, the Suzaku data provide continuous coverage of the source for more than
twice the XMM-Newton observation.

Fig. 1 shows the 0.4--10 keV light curve of NGC4395 obtained from the
three XIS cameras. The resolution is 100 s and the count rate is 
the sum of the three XISs with background correction. Large amplitude
flux variations are clearly seen. Flux doubling in 100-s and shorter intervals is
common (Fig. 2) and consistent with the characteristic timescale of
the variability of a few 100 s (Vaughan et al 2005).

While the fractional variability throughout the observation is $F_{\rm
  var} = 0.60\pm 0.06$ (see Vaughan et al 2005 for a comparison with
the XMM-Newton results), the variability over a shorter duration can
be examined for this rapidly variable object. Fig. 3 presents the
variation in fractional variability amplitude for 5 ks intervals in
a time sequence obtained from the light curve in Fig. 1. A few
intervals of excess variability correspond to periods of strong,
very short-term flaring within 1000-s or shorter duration, notably in the
hard X-ray band (see Fig. 4). The typical (e.g., median) source
flux during those periods is low. According to the correlation
between total rms and flux that is generally found in black hole systems
(Uttley \& McHardy 2001), the quantities measured here are on average expected to
remain constant. The measured values are clearly variable
as expected from the random realisation of a source with a
red-noise spectrum in flux variation (e.g., O'Neill et al 2006). The
mean and standard deviation of $F_{\rm var}$ in Fig. 3 is 0.32 and
0.13, respectively, as indicated in the figure. Nominal fluctuation
of $F_{\rm var}$ could be represented by the standard deviation. All
measurements apart from one are within $3\sigma$ fluctuation. The largest
variability recorded in the 80 ks region (Fig. 2) exceeds $3\sigma$.

\begin{figure}
\centerline{\includegraphics[width=0.2\textwidth,angle=0]{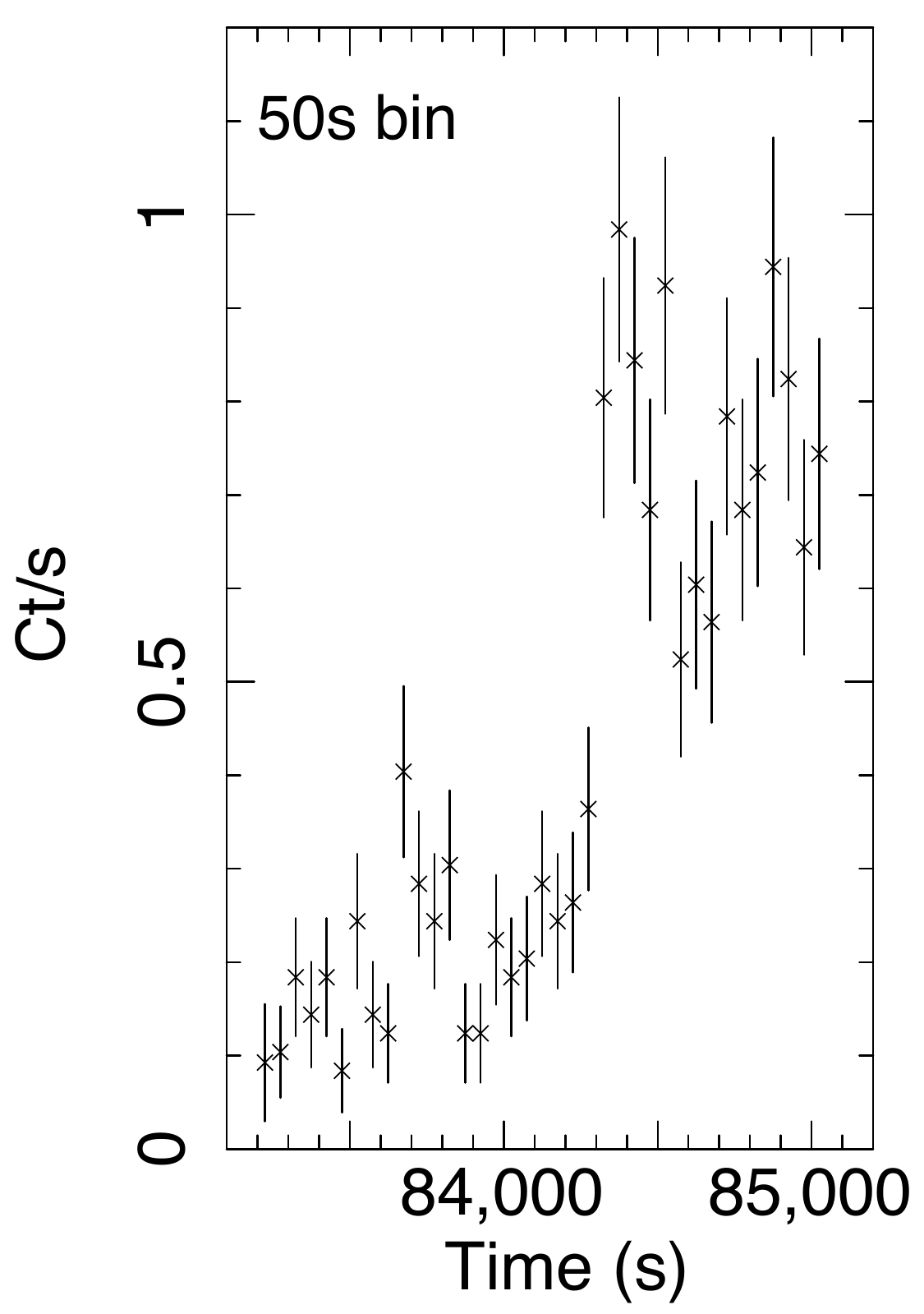}}
\caption{The blow-up of part of the light curve in Fig. 1 around 84 ks where a
  rapid and large-amplitude flare occurs. The resolution of the light curve here
  is 50 s. This strong flare causes the largest $F_{\rm var}$ in Fig. 3.}
\end{figure}

\begin{figure}
\includegraphics[width=0.43\textwidth,angle=0]{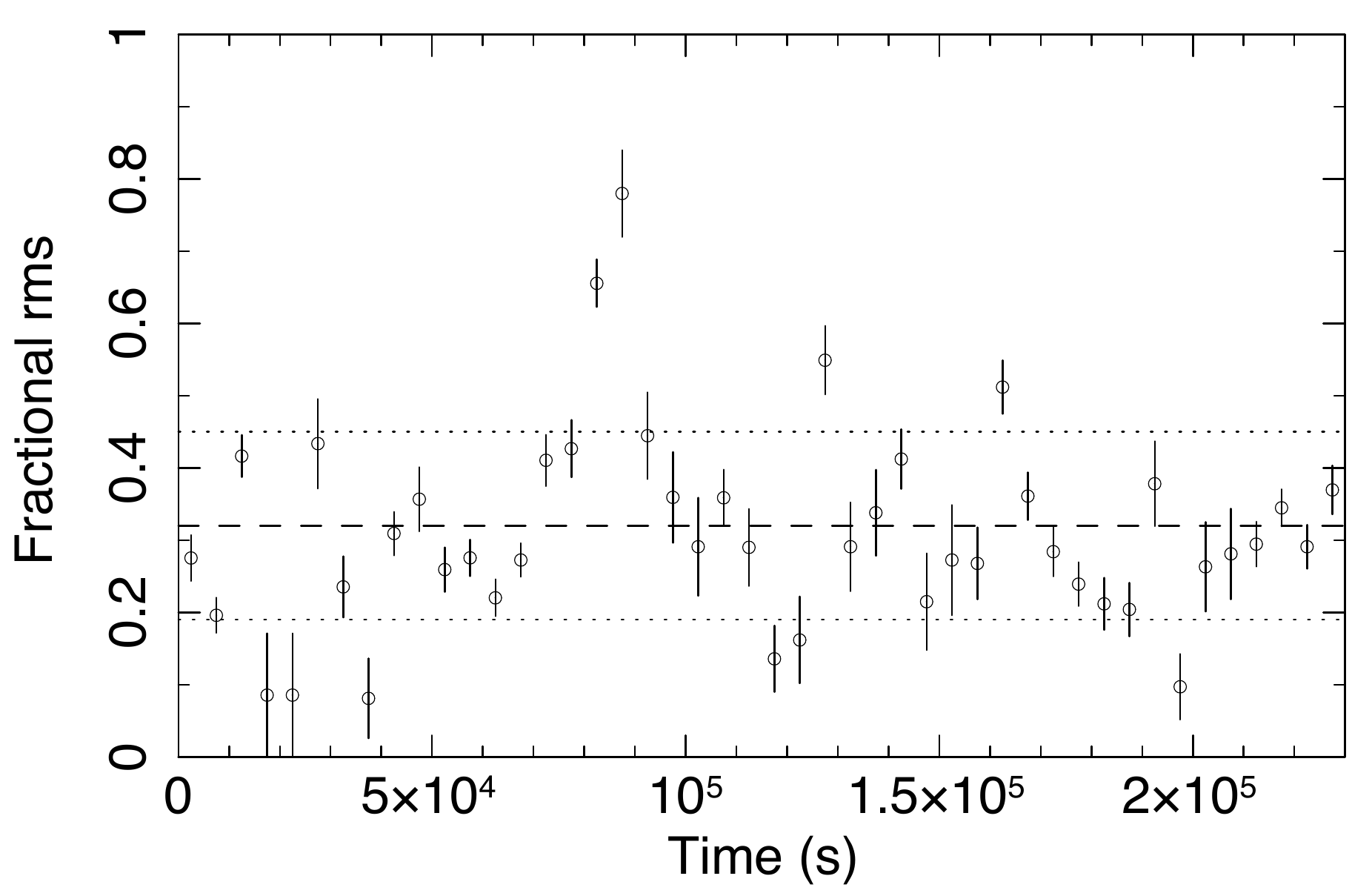}
  \caption{ The variation in the fractional variability amplitude with
    5-ks intervals, obtained from the light curve in Fig. 1. The mean
    value (0.32) and the standard deviation are marked by the
    dashed line and dotted lines, respectively. }
\end{figure}

% Fig. 4   norm lc
\begin{figure}
\includegraphics[width=0.47\textwidth,angle=0]{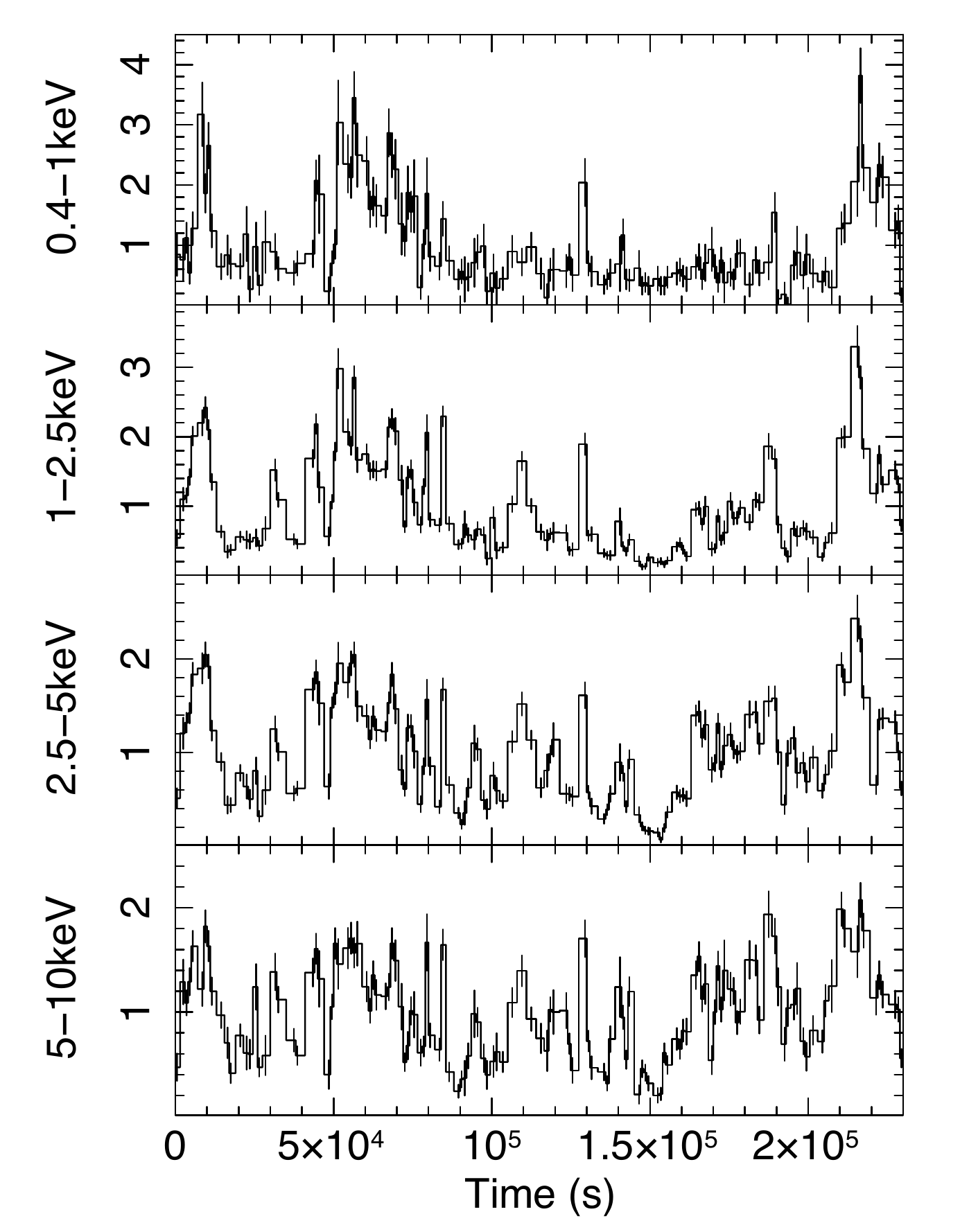}
\caption{ The X-ray light curves in four energy bands 0.4-1 keV,
  1-2.5 keV, 2.5-5 keV, and 5-10 keV (from top to bottom) normalized
  to the mean count rate. The time resolution is 1 ks. The mean count
  rates in the respective bands are 0.019, 0.109, 0.118, and 0.079
  count s$^{-1}$.  The y-axis range for each
  panel is adjusted for visual clarity and varies between the
  panels. The relative amplitude of the respective light curves can be
  quantified in Fig. 5. }
\end{figure}

% Fig. 5   cr histogram for vsoft and hard bands lc
\begin{figure}
\hbox{\includegraphics[width=0.24\textwidth,angle=0]{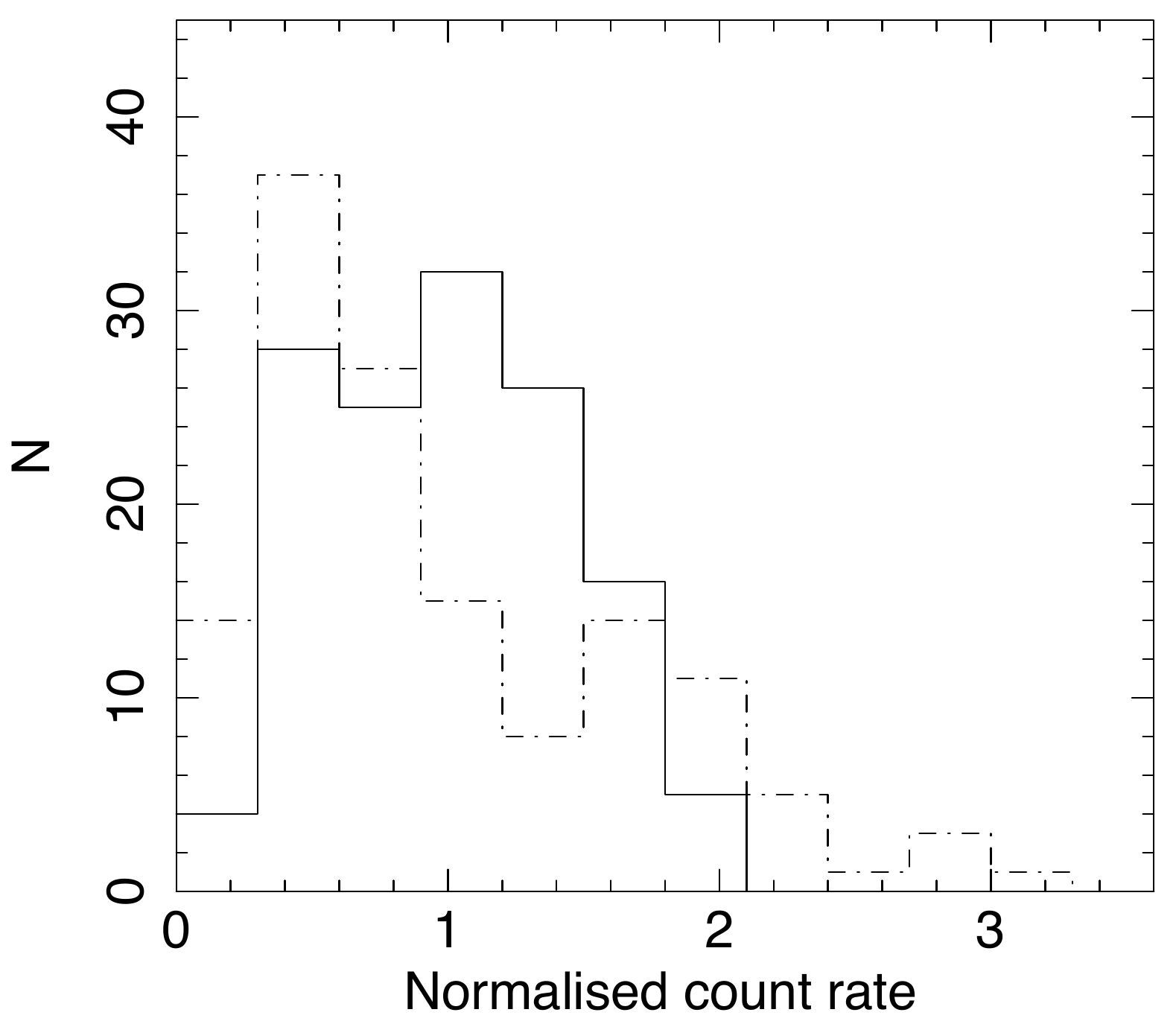}
\includegraphics[width=0.24\textwidth,angle=0]{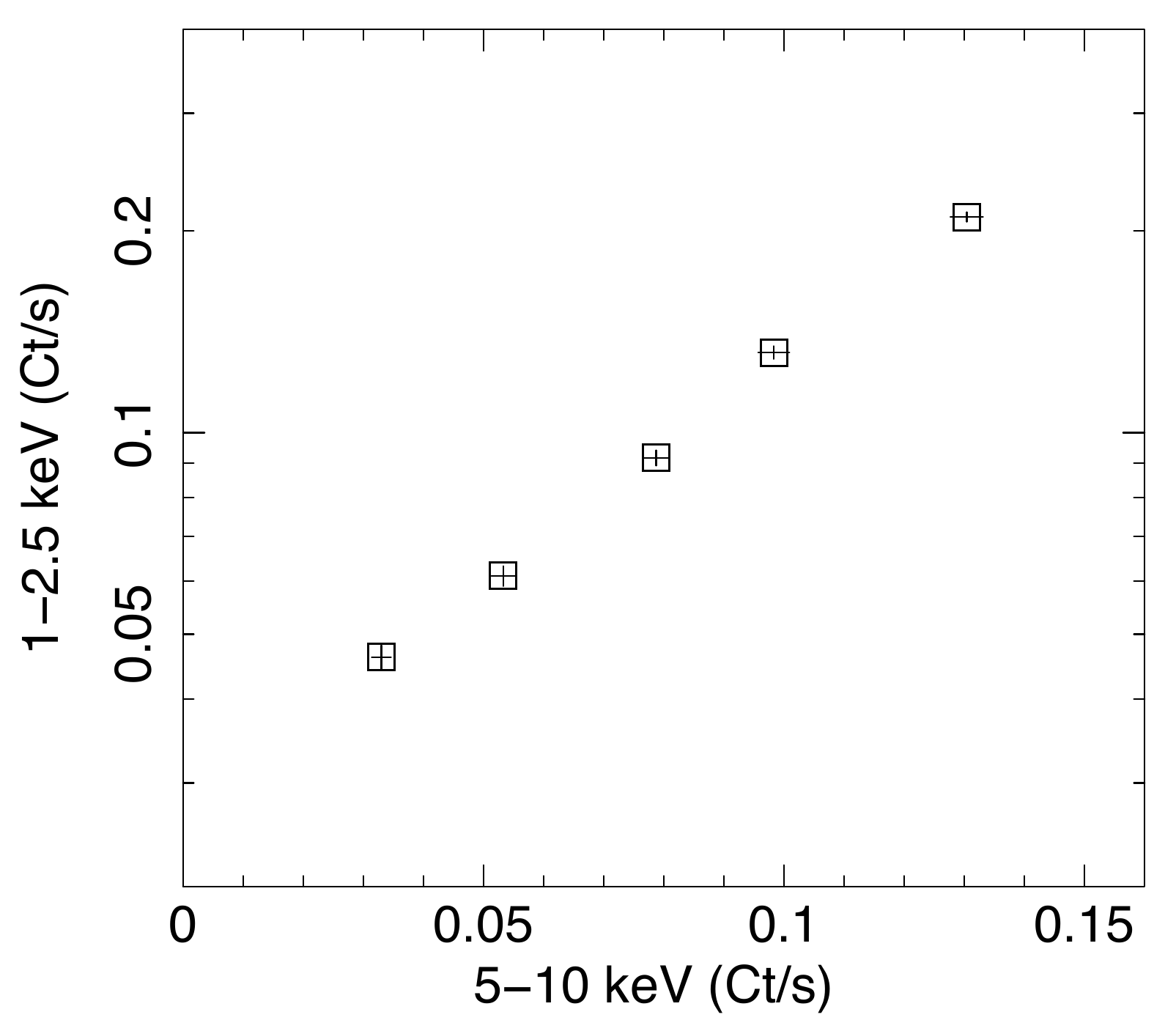}}
\caption{Left: The distribution of the normalized count rates in the
  1-2.5 keV (dash-dotted line) and 5-10 keV (solid line) bands in
  Fig. 3. It is clear that the distribution in the soft-band light
  curve is strongly skewed and well matched by a log-normal shape, while that
  for the hard band is close to being symmetric (i.e., a Gaussian). Right:
  The correlation diagram between the count rate in the 5-10 keV and 1-2.5
  keV band. The data have been incrementally rebinned according to the
  5-10 keV count rate, and the five bins contain 27, 27, 27, 27, and 28
  measurements in ascending order. Note that the y-axis (1-2.5 keV count
  rate) is in a logarithmic scale while the x-axis (5-10 keV count rate)
  is in linear scale. }
\end{figure}

% Fig. 6 rms spectrum
\begin{figure}
\includegraphics[width=0.43\textwidth,angle=0]{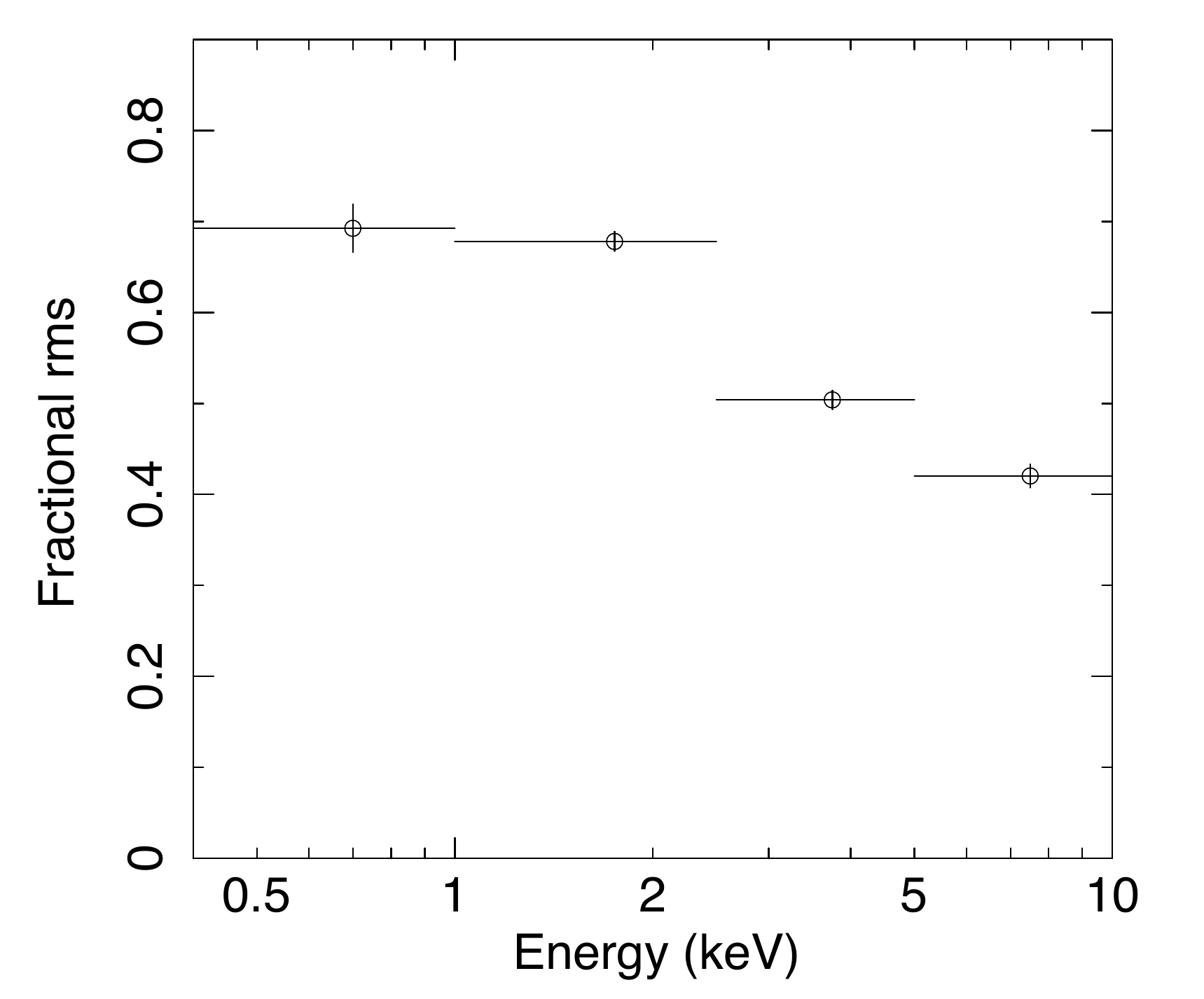}
\caption{ The fractional rms variability amplitude ($F_{\rm var}$) as
  a function of energy, obtained from the light curves in 1 ks
  resolution (see Fig. 3).}
\end{figure}

The shape of the light curve depends on the chosen energy band. Light curves
with 1 ks resolution in the four bands 0.4-1 keV, 1-2.5 keV, 2.5-5
keV, and 5-10 keV are produced and normalized to their respective mean
count rates (Fig. 4). The distribution of the source count rate is
far more skewed in the soft band than in the hard. This is
demonstrated in Fig. 5, in which only the 1-2.5 keV and 5-10 keV light
curves are displayed for clarity. The distribution in the 1-2.5 keV curve
is well matched to log-normal, while that for the 5-10 keV band is
close to being a Gaussian. The (rebinned) correlation diagram on the right
also demonstrates that 1-2.5 keV flux is correlated with 5-10 keV flux
exponentially rather than linearly, reflecting the difference in light-curve shape mentioned above. The 0.4-1 keV band distribution 
is even more skewed than at 1-2.5 keV, but the low count rate
means that distortion by the Poisson noise could be responsible for the
extra skewness.

It is known that the light curves of accreting compact objects
generally exhibit a log-normal distribution (e.g., Uttley, McHardy \&
Vaughan 2005), which is skewed towards the brighter end, as seen in
the 1-2.5 keV light curve. We note that where a light curve has a
log-normal nature, the asymmetry of the count distribution will be reduced
as the fractional variability becomes significantly small (see
simulated light curves with the fractional variability of 17, 41, and
64 per cent in Fig. 4 in Uttley et al 2005). However, in NGC4395, the
5-10 keV band variability remains strong, and the reduction in the
fractional variability (from 70 per cent in the soft band to 40 per
cent in the 5-10 keV, Fig. 6) may not be large enough to account for
the change observed. As described below, variability in the ionized
absorber could alter the shape of the distribution. Another cautionary
note is whether the observation is sufficiently long to cover the
timescale on which the stationarity of the time series is achieved,
unlike measurements for Galactic binaries.

The log-normal shape of a distribution is expected in multiplicative
products of random events, and, in the context of black hole systems, a
model in which fluctuations in the accretion rate are amplified as
they propagate inwards has been proposed (e.g., Lyubarskii 1997). In
contrast, the symmetric shape of the hard-band light curve may be
unusual. A superposition of random shots is expected to produce a
Gaussian distribution, as seen in the hard X-ray light curve. While
the soft and hard X-ray emission in NGC4395 appear to behave
independently on some occasions (Fig. 4), they are statistically
correlated (Fig. 5 right). The correlation occurs in a way that the
soft X-ray variability is stretched to higher fluxes, i.e., in an
exponential manner relative to the hard X-ray (Fig. 5 right). Seyfert
galaxies generally shows an approximately linear correlation in two
bands with an offset, which is attributed to a constant component such
as reflection, while a power-law relation, which is generally expected
(Uttley et al 2005), is more successful in reproducing data for the highly variable NGC4051
(Taylor, Uttley \& McHardy 2003).

We note that in Fig. 4, the two relatively broad, 
intervals containing high flux at 50-90 ks and 160-190 ks have different energy dependences
(also see Section 3.4). Very short-period/rapid flickering at
the 80-90 ks intervals (where the largest variability
is recorded in Fig. 3; see also Fig. 2) appears more pronounced at higher
energies. The flaring at 90-100 ks is only clear for the two harder
bands.

The overall variability amplitude is larger at
lower energies, as can be seen in Fig. 6. The fractional
variability is similar within the two low energy bands at $\sim 70$ per
cent, and declines towards higher energies to $\sim 40$ per cent
in the 5-10 keV band.

\subsection{PIN band variability}

% PIN lc
\begin{figure}
\includegraphics[width=0.43\textwidth,angle=0]{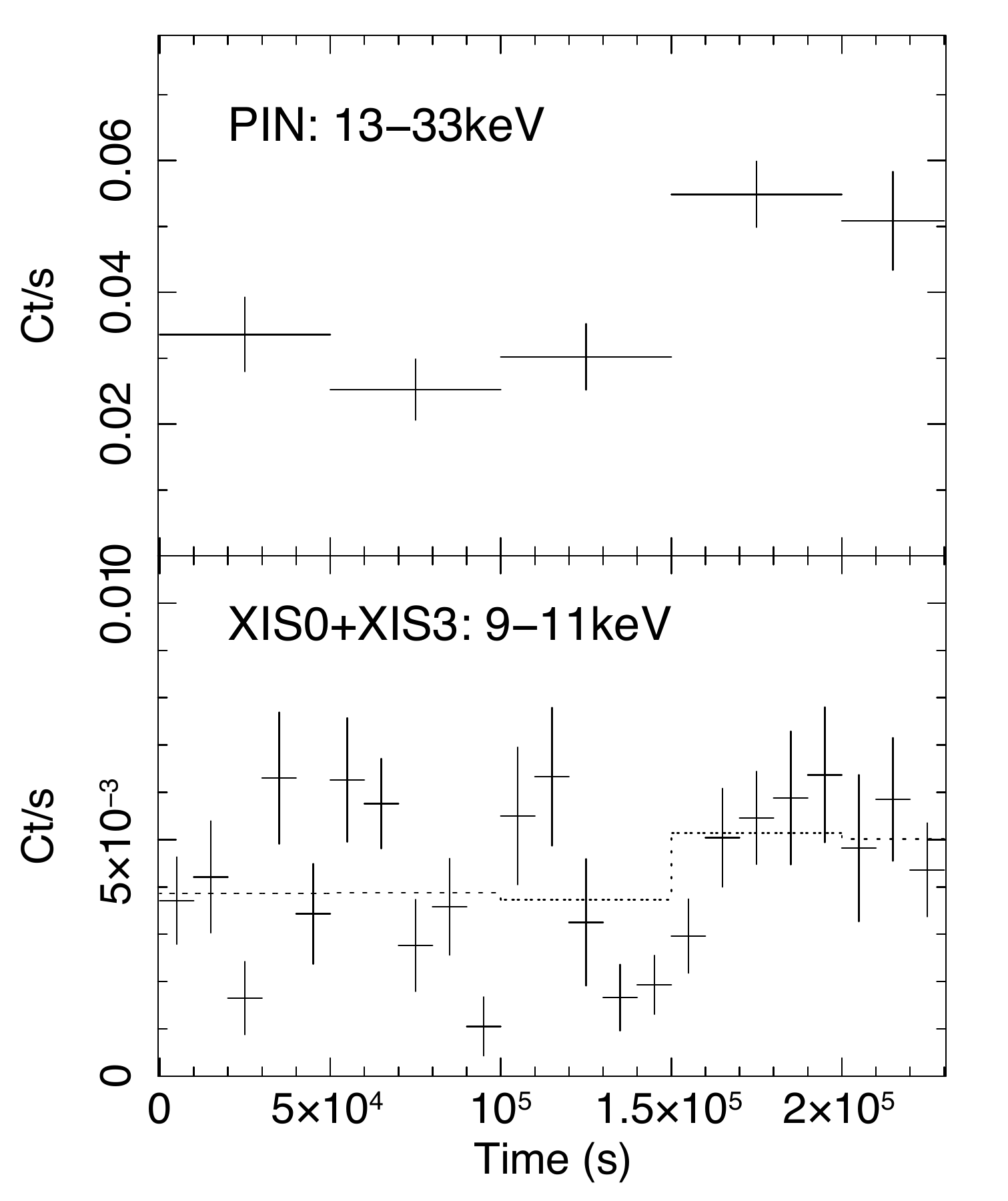}
\caption{The light curves of NGC4395 in the 13-33 keV band from the
  HXD-PIN (upper panel) and the 9-11 keV band from the
  XIS0 and the XIS3 (bottom panel). The dotted line indicates the mean
  counts over the same interval as that of the HXD-PIN light curve.}
\end{figure}

The entire observation is divided into 5 time-segments: the first four
segments are of 50 ks duration and the last of 30 ks. The 13-33 keV count
rates from the PIN for these five intervals are plotted in Fig 7. The
error bars in the count rate are statistical only. There appears to be
a flux increase by $\sim 0.02$ ct s$^{-1}$ in the final 80 ks of the
observation. In this energy band, 5 per cent of the non X-ray
background of the PIN is about 0.018 ct s$^{-1}$, which is the
conservative estimate of the uncertainty given the exposure of the
individual data points. The possible flux variation is
comparable to this, but the reality of the apparent flux variation
is uncertain. As a comparison, the light curve for the 9-11 keV band,
which was obtained by the XIS0 and XIS3 (the signal from the FI-CCD of the
XIS1 is very noisy in this band), is also shown in Fig. 7. Flux
variability of $\simeq 30\pm 5$ per cent is still present at the
highest energy end of the XIS bandpass when viewed on the 5 ks time
resolution. On the 50 ks timescale, the amplitude is lower, but a
similar variability pattern between the two light curves indicates
that the hard X-ray variability in the PIN band may be real.

\subsection{The time-averaged energy spectrum}

% XIS flux density spec
\begin{figure}
\includegraphics[width=0.45\textwidth,angle=0]{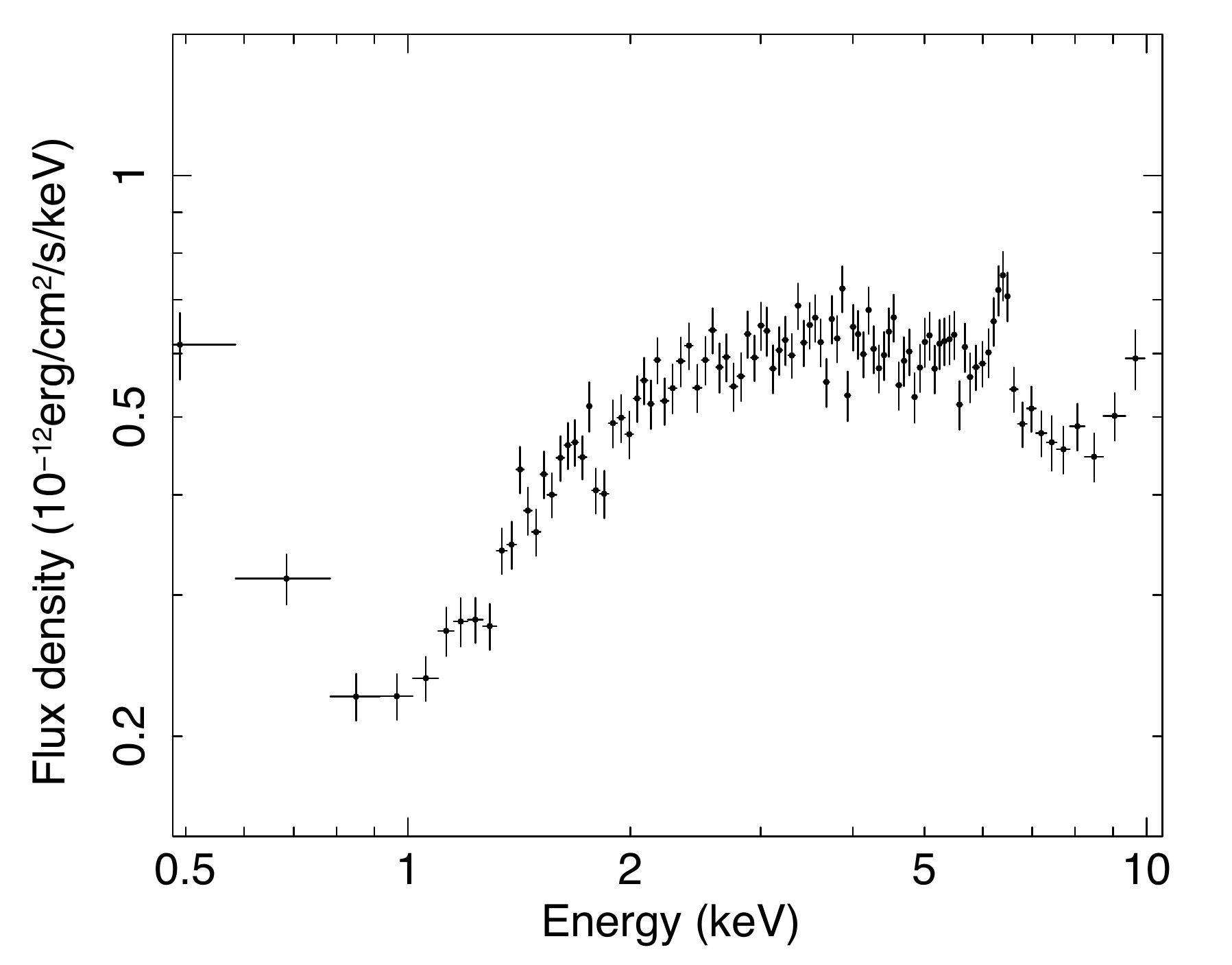}
\caption{ The 0.5-10 keV spectrum of NGC4395 observed with the Suzaku
  XIS. The data are obtained from the XIS0 and XIS3, and plotted in
  units of flux density after correcting for the detector
  response. This correction may have some uncertainty but is accurate
  enough to show the major spectral features in the spectrum. }
\end{figure}

% XIS + PIN spec
\begin{figure}
\includegraphics[width=0.45\textwidth,angle=0]{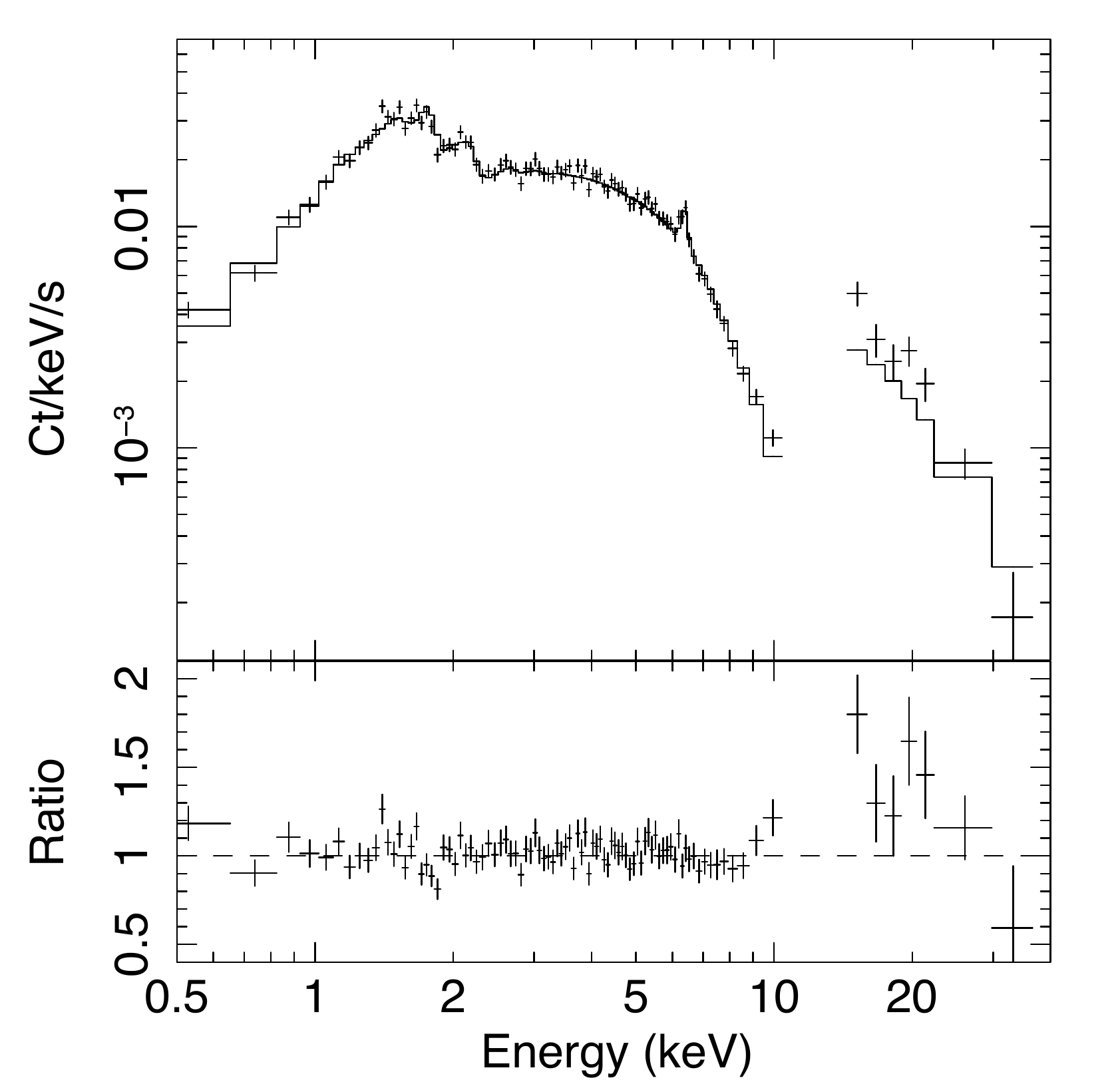}
\caption{ The 0.5-35 keV spectral data of NGC4395 obtained with the XIS
  and HXD-PIN. The solid-line histogram shows the model of a
  power law with the Fe K line modified by the warm absorber folded
  through the detector responses, which fits the XIS data. The
  cross-calibration constant is assumed to be 1:1.18 (XIS:PIN).}
\end{figure}

\subsubsection{The XIS data}

A simple power law is  unacceptable in describing the full-band
data mainly because of the strong absorption at low energies 
(Fig. 8).  The best-fit model photon index $\Gamma\sim 0.7$ indicates that the
overall spectrum is unusually hard. Because of the strong absorption and
hard spectrum, the spectral analysis here focuses mainly on the
two FI CCD cameras, XIS0 and XIS3.

Introducing a warm absorber and a Fe K emission-line provides a good
description of the 0.4-10 keV band spectrum (Fig. 9). The power-law
continuum has a photon-index of $\Gamma = 1.40^{+0.05}_{-0.04}$. One of
Tim Kallman's XSTAR grid tables ({\tt grid25}, which is the
photoionization grid computed by XSTAR with a power-law ionizing
source and the turbulent velocity of the ionized gas of 200 km
s$^{-1}$)\footnote{http://heasarc.gsfc.nasa.gov/docs/software/xstar/xstar.html}
is used to model the warm absorber. The absorbing column of the warm
absorber is $N_{\rm H,W}=2.1^{+0.3}_{-0.2}\times 10^{22}$\psqcm, and the
ionization parameter log $\xi = 1.53^{+0.6}_{-0.5}$ (erg cm
s$^{-1}$). A small cold absorption of \nH $=5.4\times 10^{20}$ \psqcm,
in addition to the Galactic column (\nH = $1.9\times 10^{20}$\psqcm,
for the LAB map by Kalberla et al 2005) is also required. The Fe K line,
modelled with a Gaussian profile, has a centroid energy at $6.36\pm 0.04$
keV, a line width $\sigma = 0.09\pm 0.05$ keV (or FWHM$\sim 10,000$
km s$^{-1}$), and line intensity $6.4^{+2.2}_{-1.8}\times 10^{-6}$
\phpspsqcm, which corresponds to an equivalent width of $EW =
115^{+40}_{-32}$ eV. The quality of fit ($\chi^2 = 406.0$ for
407 degrees of freedom) is good (Table 1). We note that there is a
hint of systematic hardening above 9 keV (see Fig. 8 and the residual
plot in Fig. 9). However, the data in this range have small
statistical weight in the full-band spectrum and the deviation does
not have an impact on the quality of the overall fit.

The observed flux is $5.1\times 10^{-13}$ \ergpspsqcm\ in the 0.5-2 keV
band and $4.4\times 10^{-12}$ \ergpspsqcm\ in the 2-10 keV band. The
2-10 keV flux is similar to that during the long ASCA observation
($4.7\times 10^{-12}$\ergpspsqcm, Shih et al 2002) and slightly ($\sim
20$ per cent) lower than that during the long XMM-Newton observation
($5.6\times 10^{-12}$\ergpspsqcm, Vaughan et al 2005).

\begin{table*}
\begin{center}
  \caption{Spectral fit to the Suzaku XIS time-averaged spectrum.}
    \begin{tabular}{ccccccccc}
Data & $\Gamma$ & \nH & $N_{\rm H,W}$ & $\xi$ &
$E_{\rm FeK}$ & $\sigma$ & $I_{\rm FeK}$ & $\chi^2$/dof \\
& & \psqcm & \psqcm
& erg\thinspace s$^{-1}$\thinspace cm
&keV & keV & \phpspsqcm & \\[5pt]
Total & $1.40^{+0.05}_{-0.04}$ & $5.4^{+2.8}_{-3.0}\times 10^{20}$
& $2.1^{+0.3}_{-0.2}\times 10^{22}$ & $34^{+5}_{-4}$ & $6.36^{+0.04}_{-0.04}$ & $0.09^{+0.05}_{-0.05}$ & $6.4^{+2.2}_{-1.8}\times 10^{-6}$ & 406.0/407 \\[5pt]
%A & $1.56^{+0.08}_{-0.08}$ & --- & $2.0^{+0.3}_{-0.3}\times 10^{22}$
%& --- & --- & --- & $7.7^{+1.6}_{-5.0}\times 10^{-6}$ & 386.2/343 \\
%B & --- & --- & $4.6^{+0.7}_{-0.7}\times 10^{22}$ & --- & ---
%& --- & --- & --- \\
\end{tabular}
\begin{list}{}{}
\item[Note:] The spectrum was obtained by combining the XIS0 and XIS3
  data. The model consists of a power law with a Gaussian for the Fe
  K$\alpha$ line, modified by warm and cold absorption. The cold
  absorption column is denoted by \nH, which is in excess of the
  Galactic column of $1.85\times 10^{20}$\psqcm. Tim Kallman's XSTAR
  grid tables (``grid 25'') with the turbulent velocity of 200 km
  s$^{-1}$ is adopted for the warm absorber modelling.
\end{list}
\end{center}
\end{table*}

\subsubsection{PIN data}

The 15-35 keV spectrum measured with the HXD-PIN appears to be much
steeper than that below 10 keV, albeit with a large
uncertainty. Fitting a power law provides a reasonable fit to the data
with a photon index of $\Gamma = 2.2\pm 0.5$ ($\chi^2 =14.5$ for the 8
degrees of freedom). The inclusion of any spectral curvature or break
does not improve the fit. The 20-40 keV flux, estimated by the simple
power-law model, is $5.9\times 10^{-12}$\ergpspsqcm, which is similar
to those measured previously by the INTEGRAL IBIS (Beckman et al 2006;
Bird et al 2007). 

%The second INTEGRAL AGN catalogue (Beckmann et al
%2009), using the data up to Feb 2007, gives a higher 20-40 keV flux of
%$1.5\times 10^{-11}$ \ergpspsqcm.

The broad-band spectrum is examined by combining the XIS and the PIN
spectrum. Since the observation efficiency of the PIN is slightly
lower than that of the XIS, good time intervals for the PIN are also used
to produce the XIS spectrum analysed here ($\sim 85$ per cent of the
original XIS exposure).

When extrapolating the best-fit model of the XIS spectrum (Fig. 9),
the lower end of the PIN data are well above the extrapolation. The
cross-calibration constant of
1.18\footnote{ftp://legacy.gsfc.nasa.gov/suzaku/doc/xrt/suzakumemo-2008-06.pdf}
between the PIN and the XIS is adopted, which was measured for the
Crab nebula at the HXD nominal pointing position. Provided this cross
calibration is reliable, a spectral hump in the 10-25 keV range is
required, which may be supported by the possible spectral hardening
seen at the highest energy end of the XIS data (Figs. 8 and 9). However, we caution that the uncertainty in
the PIN background estimate may significantly influence whether we detect a spectral hump for this faint source. The source
fraction of the total PIN counts in the energy band of interest is
only $\sim 9$ per cent. $\pm 3$ per cent uncertainty in the background
estimate translates into the cross-calibration constant in the range of
0.8-1.6 being allowed. A close match between the extrapolated
model and the 15-20 keV PIN data can indeed be obtained when the
cross-calibration constant is increased to $\sim 1.5 $.

\subsection{Spectral variations}

As inferred by the study of light curves in various energy bands, the
spectral shape of the X-ray source in NGC4395 is also
variable. Furthermore, the spectral variability does not appear to
correlate with the source brightness (see Section 3.1).

The throughput of the Suzaku XIS is insufficient to track the
spectral changes on the time-scale of the most rapid flux variability
(of the order of 100 s). Instead, the spectral variability was
investigated on a timescale of 5-ks using X-ray colours. The four
energy bands 0.4-1 keV, 1-2.5 keV, 2.5-5 keV, and 5-10 keV were used
in the X-ray colour analysis. The X-ray colour is defined to be
HR = (H-S)/(H+S), where H and S are the background-corrected count
rates collected from the three XIS cameras in two bands of choice and
H is always of the higher energy. Four HR (HR1, HR2, HR3, and HR4)
are derived as shown in Table 2. 

The X-ray colour measurements are
plotted in Fig. 10, along with the full-band light curve. Four
correlation-diagrams between the X-ray colours and the source
brightness are presented in Fig. 11. Overall, there is no clear
correlation between the spectral variability and source flux, although
some systematic trends can be found: 

1) HR values increase towards lower flux down to the 0.4-10 keV count
rate of $\sim 0.2$ ct s$^{-1}$, except for HR1, 2) Below this flux,
HR drops, except for HR3. The former trend is consistent with spectral
pivoting, i.e., the spectral slope change as a function of source
brightness. However, the turnover of the HR value in the low flux
range suggests further complication to a simple spectral
pivoting. Below 1 keV, there is an emission component from
photoionized gas, detected through XMM-Newton RGS measurement of OVII
(Bianchi \& Guaunazzi 2007), which is detached from the variable
nuclear emission and expected to be stable. This could partly explain the HR turnover, namely in HR1 and HR4, as the stable
component becomes visible when the nuclear flux is low.
%% 

% Table of HR

\begin{table}
\begin{center}
  \caption{The definition of the four X-ray colours.}
\begin{tabular}{lccc}
HR & S & H & $\chi^2$/dof \\
& keV & keV & \\[5pt]
HR1 & 0.4-1 & 1-2.5 & 42.2/45 \\
HR2 & 1-2.5 & 2.5-5 & 323/45 \\
HR3 & 2.5-5 & 5-10 & 115/45 \\
HR4 & 0.4-1 & 5-10 & 117/45 \\
\end{tabular}
\begin{list}{}{}
\item[Note:] Each colour indicator (HR) is defined by (H-S)/(H+S), where H and
    S are count rates measured in the two energy bands given
    above. $\chi^2$ values for a constant hypothesis are also
    included.
\end{list}
\end{center}
\end{table}

% Fig. HR curves

\begin{figure}
\centerline{\includegraphics[width=0.45\textwidth,angle=0]{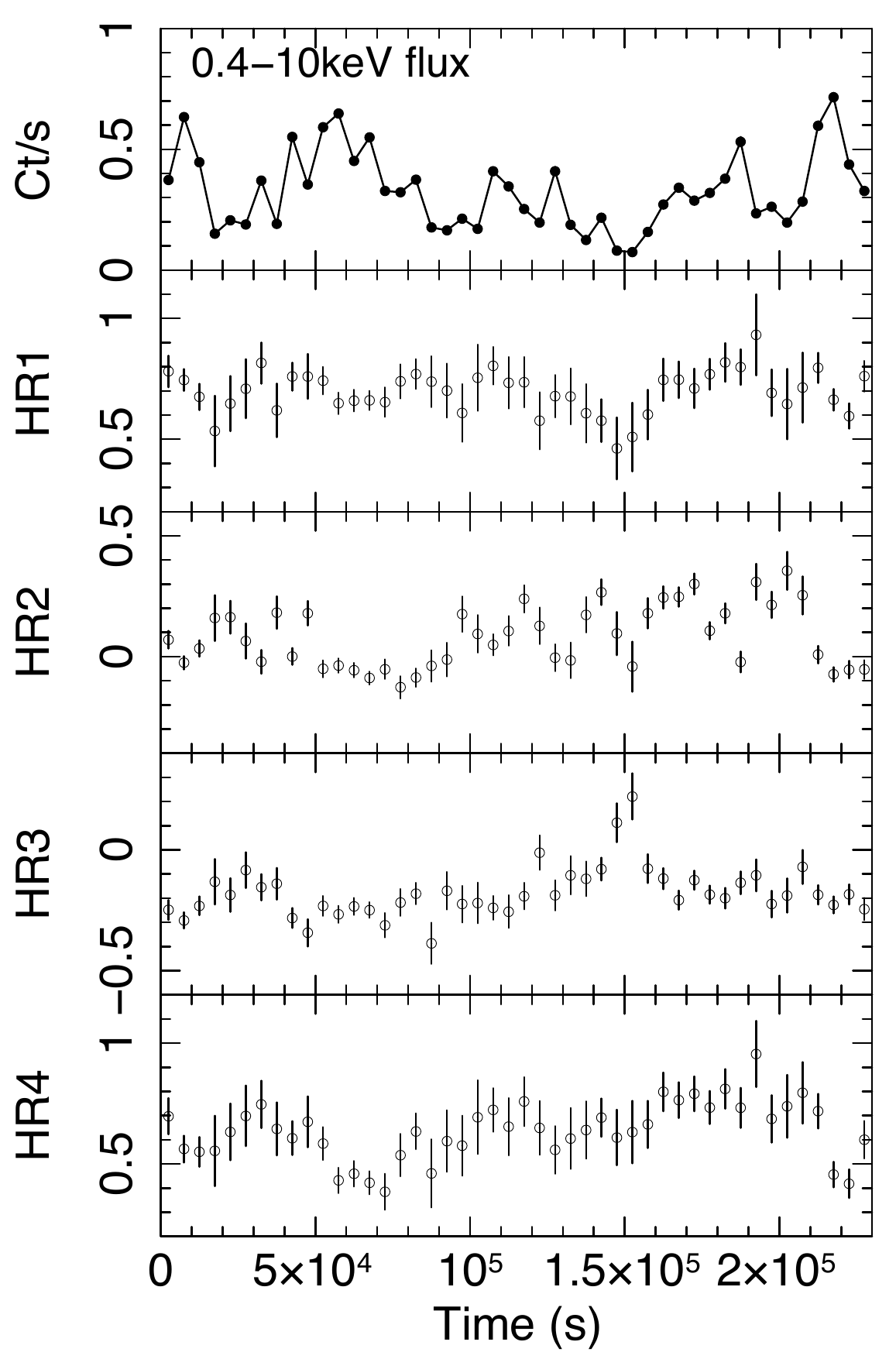}}
\caption{The temporal variations in the 0.4-10 keV count rate and the four
  X-ray colours (HR1, HR2, HR3, and HR4) at 5 ks resolution (from top to
  bottom). }
\end{figure}

% Fig HR vs cps x3
\begin{figure}
\includegraphics[width=0.47\textwidth,angle=0]{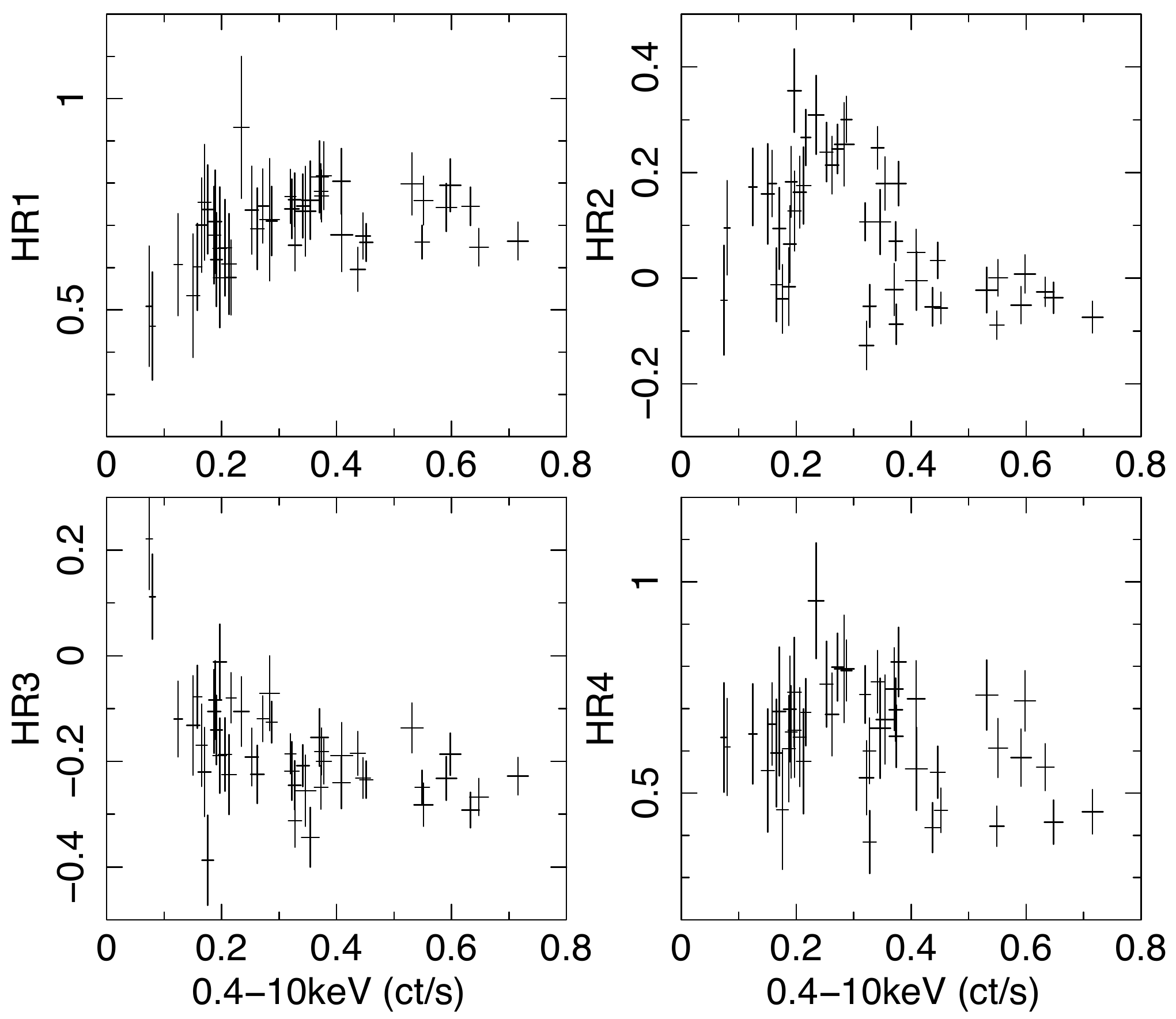}
\caption{ The three X-ray colour indicators HR1, HR2, HR3, and HR4 (see
  Table 2) plotted against the 0.4-10 keV count rate. }
\end{figure}

% Table six spectral fits

\begin{table}
\begin{center}
  \caption{The XIS 1-10 keV spectra of the six time-intervals.}
\begin{tabular}{lcccc}
Interval & $\Gamma $ & log $\xi^{\mathrm{a}}$ & $N_{\rm W}^{\mathrm{b}}$ & $F_{\rm 2-10}^{\mathrm{c}}$\\[5pt]
1 & $1.50\pm 0.07$ & $1.6\pm 0.1$ & $1.6\pm 0.2$ & 4.8 \\
2 & $1.67\pm 0.06$ & $1.7\pm 0.1$ & $1.2\pm 0.2$ & 5.2 \\
3 & $1.32\pm 0.08$ & $1.3\pm 0.1$ & $2.0\pm 0.2$ & 3.3 \\
4 & $1.08\pm 0.06$ & $1.0\pm 0.1$ & $2.4\pm 0.2$ & 2.6 \\
5 & $1.16\pm 0.06$ & $1.0\pm 0.1$ & $2.4\pm 0.2$ & 4.9 \\
6 & $1.58\pm 0.06$ & $1.6\pm 0.1$ & $1.3\pm 0.2$ & 5.6 \\[5pt]
$\Gamma $ & --- & $1.47\pm 0.04$ & $1.40\pm 0.04$ \\
log $\xi^{\mathrm{a}}$ & $1.5\pm 0.1$ & --- & $1.2\pm 0.1$ \\
$N_{\rm W}^{\mathrm{b}}$ & $2.2\pm 0.3$ & $2.2\pm 0.1$ & --- \\[5pt]
$\chi^2$/dof & 671.8/676 & 707.2/676 & 698.1/676 \\
\end{tabular}
\begin{list}{}{}
\item[Note:] Results of fitting a single warm absorber model to the 1-10 keV XIS spectra from the six time-intervals are presented. One of the three
    parameters, $\Gamma $, $\xi $, and $N_{\rm W}$, is a free
    parameter when fitting the six spectra, besides the normalizations of
    the power-law continuum. The other two are tied together between the
    spectra and reported below. Observed 2-10 keV flux is also given.
\item[$^{\mathrm {a}}$] In unit of erg s$^{-1}$ cm.
\item[$^{\mathrm {b}}$] In unit of $10^{22}$ cm$^{-2}$.
\item[$^{\mathrm {b}}$] The 2-10 keV flux in unit of $10^{-12}$ erg cm$^{-2}$ s$^{-1}$.
 \end{list}
\end{center}
\end{table}

% Table 4   2nd & 5th intervals
\begin{table}
\begin{center}
  \caption{The XIS spectra fromthe second and fifth intervals.}
\begin{tabular}{lcccc}
Interval & $\Gamma $ & log $\xi^{\mathrm{a}}$ & $N_{\rm W}^{\mathrm{b}}$ & \\[5pt]
2 & $1.75\pm 0.07$ & $1.6\pm 0.5$ & $1.1\pm 0.2$ & \\
5 & $1.23\pm 0.07$ & $1.0\pm 0.1$ & $2.4\pm 0.3$ & \\[5pt]
$\Gamma $ & --- & $1.56\pm 0.09$ & $1.47\pm 0.09$ & \\
log $\xi^{\mathrm{a}}$ & $1.4\pm 0.3$ & --- & $1.1\pm 0.2$ & \\
$N_{\rm W}^{\mathrm{b}}$ & $2.2\pm 0.5$ & $2.3\pm 0.3$ & --- & \\[5pt]
$\chi^2$/dof & 281.8/309 & 293.2/309 & 283.4/309 & \\[10pt]

Interval & $\Gamma $ & log $\xi^{\mathrm{a}}$ & $N_{\rm W}^{\mathrm{b}}$ & $\chi^2$/dof \\[5pt]
2 & $1.65\pm 0.08$ & $1.5\pm 0.2$ & $2.2\pm 0.4$ & 275.0/308 \\
5 & $1.34\pm 0.10$ & $1.0\pm 0.2$ & as above & \\[5pt]
2 & $1.66\pm 0.09$ & $1.3\pm 0.2$ & $1.8\pm 0.2$ & 275.5/308 \\
5 & $1.34\pm 0.09$ & as above & $2.3\pm 0.3$ & \\[5pt]
\end{tabular}
\begin{list}{}{}
\item[Note:] Results of joint model fitting to the XIS spectra from the
    second and fifth intervals are shown. For the spectral fitting, the procedure is
    the same as that described in Table 3. The lower part shows the
    results when the spectral slope and the warm absorber parameters
    ($N_{\rm W}$, log $\xi$) are left to vary, but either of the two
    warm absorber parameters is tied between the two spectra.
\item[$^{\mathrm {a}}$] In unit of erg s$^{-1}$ cm.
\item[$^{\mathrm {b}}$] In unit of $10^{22}$ cm$^{-2}$.
 \end{list}
\end{center}
\end{table}

In the time sequence, there is considerable flaring in the 40-80 ks interval,
where the spectrum remains soft, as shown by generally low values of
HR (Fig. 10). However, the softest period, indicated by the HR curves,
seems to last for a while ($\sim 20$ ks) after the flare peak. Similar HR behaviour may also occur during the decay of the large
flare at the end of the observation. These are seen in HR2, which
shows the most significant variation (see Table 2), and possibly in
HR4. The colour HR2 is taken from the energy range where the source count rate is the
highest, which is probably the main reason for the large significance,
but where also the warm absorber modifies the spectrum the most. In the simplest picture
of photoionization, the ionization parameter should change in response
to variations in the illuminating source and the ionic column of the
absorber should also change on the recombination timescale. The
complicated HR behaviour may thus also originate from the variability
in absorption.

In an attempt to identify the main driver of the spectral variability
in NGC4395, we divided the observation into six intervals and examined
their spectra, by applying the simple model which was used to model the time-averaged spectrum (Table 1). All the time intervals are of a 40 ks
duration each, except for the last interval, which has a 30 ks
duration. The second and fifth intervals roughly correspond to the
periods, where the spectrum is persistently soft and hard,
respectively, according to HR2.

The six spectra were then fitted jointly by restricting the choice of the model free parameters to the power-law normalisations and one parameter of choice
among photon index $\Gamma $, the ionization parameter $\xi$, and the
absorption column density $N_{\rm W}$ of the warm absorber, either of
which could be the driver of the spectral variability. To avoid
possible contamination by photoionized gas at the low flux
intervals, the data below 1 keV were excluded. The cold absorption was
fixed to the best-fit model value for the time-averaged spectrum (see Table
1), and all other parameters are tied together between the six
spectra. Results are summarised in Table 3. The highest quality of fit is
obtained when $\Gamma $ is a free parameter, suggesting that the slope
change might be the primary driver of the spectral variability. 

However, as discussed for the HR analysis, the slope change may not be
the unique cause of the spectral variability. We then took the second
and fifth intervals, which have almost identical 5-10 keV flux,
$\simeq 2.9\times 10^{-12}$ \ergpspsqcm\ (see Fig 4) but significantly
different values in HR2 (see Fig. 10). The same exercise performed for
these two intervals infers that a change in the continuum slope and
the absorption depth are comparably good solutions (see Table 4). When
either $\xi$ or $N_{\rm W}$ were allowed to vary between a pair of
intervals, in addition to $\Gamma $, the quality of the fit improved
by a similar degree. The reduction in $\chi^2$ relative to the fit,
where only $\Gamma $ is variable between the two spectra (Table 4), is
$\Delta\chi^2\approx -6$, for which a $F$-test infers 99 per cent
confidence in the one additional term. No additional improvement in
the fit was attainable when both $\xi$ and $N_{\rm W}$ were allowed to
vary. This may indicate that changes in the warm absorber, at least in
one of the two parameters, could be a secondary source of the spectral
variability. Which parameter varies of either $\xi $ or $N_{\rm W}$ in
the warm absorber is difficult to determine, because their features are
degraded at the CCD resolution. The higher value of $\xi $ for the
spectrum with higher flux is qualitatively consistent with the
photoionization of the absorbing medium.

Finally, we comment on the spectrum at the minimum of the light curve at
$\sim 150$ ks. During this flux minimum, HR1 shows spectral softening,
while HR3 shows strong hardening. The softening could be explained by
the increased relative contribution of the stable, photoionized-gas
emission to the soft band. On the other hand, the larger value of HR3 is caused by rapid flaring, which only occurs in the 5-10 keV band (see
Fig. 4). Naturally, this hard X-ray flare has an extremely flat
spectrum ($\Gamma =0.25\pm 0.15$), which makes it difficult to explain
using the absorption model. In this brief period, the variable continuum
does not switch off completely, as significant variability is still
present (Fig. 2). Therefore, explanation of the hardening in HR3,
invoking a constant component (in this case, the cold reflection from
distant matter), which would appear mainly in the 5-10 keV band, does
not apply.

%Fig. 12  soft, hard  spec

\begin{figure}
\centerline{\includegraphics[width=0.45\textwidth,angle=0]{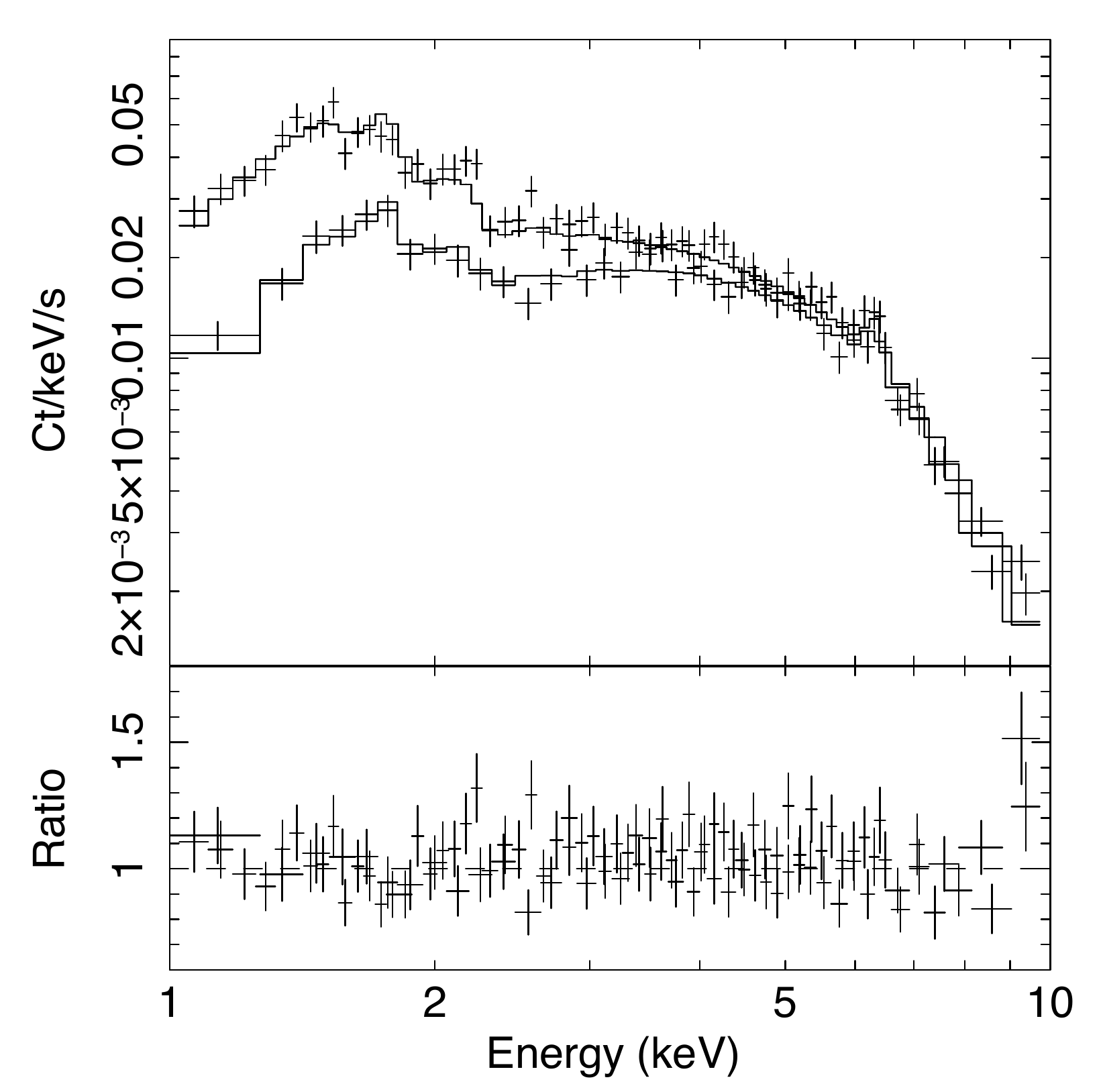}}
\caption{ The count rate spectra for the two intervals (40-80
  ks and 160-200 ks), where the 5-10 keV flux is comparable but the HR2
  shows largely different values. These two intervals correspond to
  the second and fifth intervals in Table 3, and investigated in Table
  4. The spectra are obtained by combining the data from the XIS0 and
  XIS3. The solid-line histograms indicate the best-fit model in
  which the power-law normalization and the absorption column ($N_{\rm
    W}$, see Table 4) are the only independent parameters between the
  two. The bottom panel shows the residuals in the form of the ratio
  of the data to the model for the two spectra.}
\end{figure}

%Fig. SED

\begin{figure}
\centerline{\includegraphics[width=0.47\textwidth,angle=0]{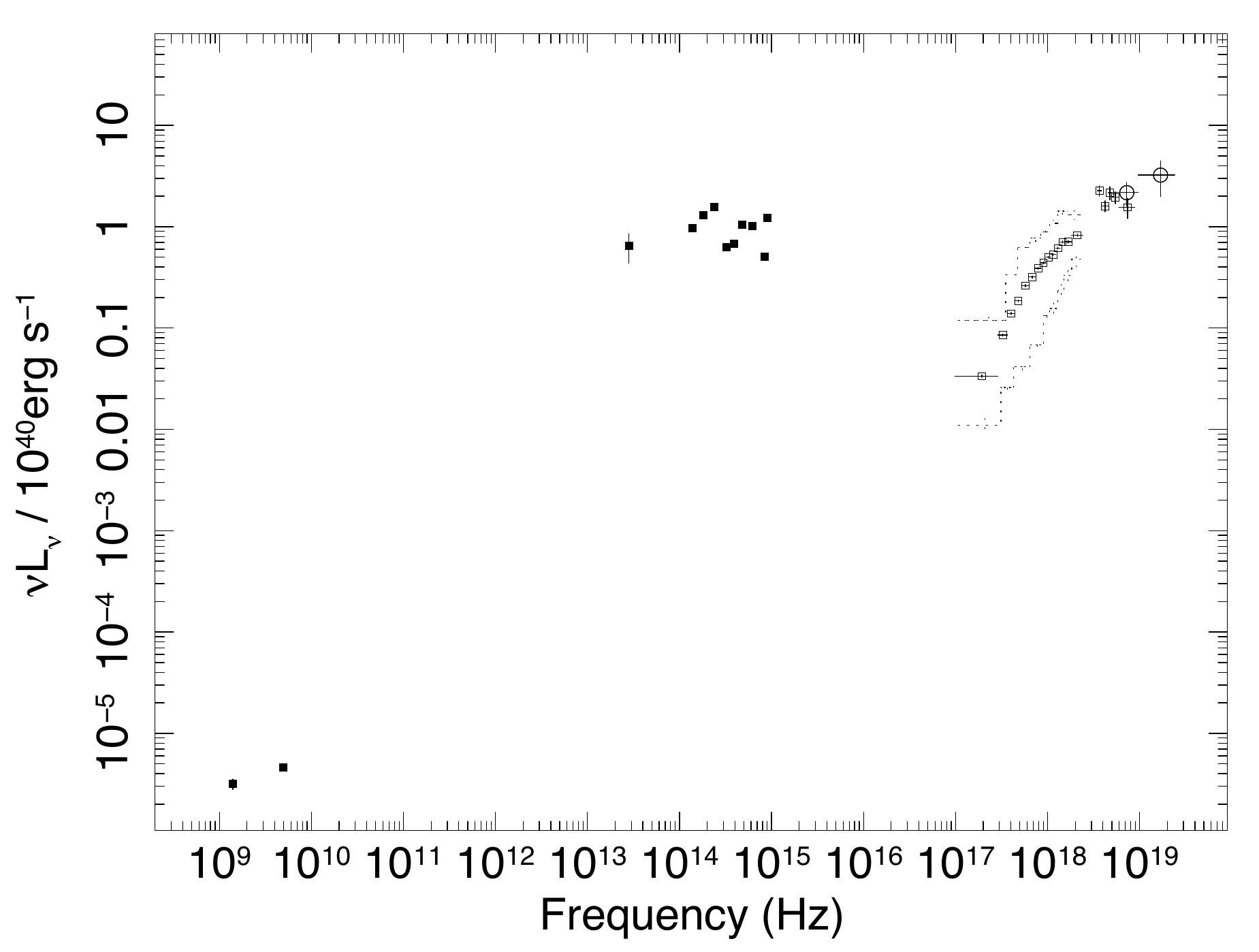}}
\caption{ The energy distribution of NGC4395 from radio to hard
  X-rays. The X-ray data (open squares) are measured with Suzaku. The
  dotted histograms indicate the spectra during the low- and
  high-flux states observed in the Suzaku observation. The 20-40 keV
  and 40-100 keV data points (open circles) are from INTEGRAL IBIS
  (Bird et al 2007). The other points are taken from NED, selecting
  photometric data from the smallest aperture: SDSS (u, g, r, i, z),
  HST/ACS (F330W filter, 0.3 arcsec aperture, Mu\~noz Martin et al
  2007), 2MASS (J, H, K$_S$, 14 arcsec aperture), MMT (N, 5.3 arcsec
  aperture, Maiolino et al 1995), VLBA (5 GHz, Nagar, Falcke \& Wilson
  2005), and FIRST (1.4 GHz, Becker, White \& Helfand 1995). IRAS fluxes
  are not included since they are believed to originate in the galaxy,
  not from the nucleus. The energy distribution of NGC4395 is unusual for
  Seyfert galaxies or quasars (e.g., Elvis et al 1994), the
  energy output at 10 keV or above being comparable to that in the
  optical band. }
\end{figure}

\section{Iron line variability}

The iron K line variability was investigated in time and also at
different continuum flux levels. Since the line is weak (Table 1), it is not
possible to track the response of the line flux to the extremely rapid
continuum variation. The XIS spectra from three time intervals {\it
  i-1} of 0-90 ks, {\it i-2} of 90-160 ks, and {\it i-3} of 160-230 ks were
extracted and the line flux measured. The intervals, {\it i-1} and
{\it i-3} contain periods of strong flaring, while {\it i-2} represents
relatively quiet periods.

%The XIS spectra were obtained from the XIS0 and XIS3. 
The spectral
data are restricted to the 3-10 keV range and the continuum and the
iron line are modelled by an absorbed power law and a Gaussian,
respectively. The line flux is not corrected for absorption.
The line fluxes in the three intervals  agree and show no
sign of variability, as shown in Table 5.

Next, the source light curve was sliced into three count-rate ranges
and the line flux from each source brightness range was measured. The
count-rate slicing was applied to the light curve in Fig. 1 (in the
0.4-10 keV band with the 100-s time resolution) with the count rate
ranges of L: $\leq 0.30$ \cps, M: 0.30-0.55 \cps, and H: $\geq 0.55$
\cps. The line flux measurements were performed in the same way as for
the time-ordered spectra. The results are presented in Table 5. There
is a hint of line flux increase in the data at the highest source
brightness, but within the uncertainty, the line flux does not show
significant variability over factor of $\sim 3$ change in average
flux. At least, a linear correlation between the continuum and line
flux can be ruled out. Table 3 also shows exposure time, the collected
counts, and the flux in the 3-10 keV band. 

%%%
Ideally, the flux slicing should be performed using the
ionizing flux, e.g., the 7-11 keV band flux responsible for
ionizing Fe. The much lower count rate in this hard band means
however that the time resolution becomes poorer. We made three
flux slices based on the 7-11 keV light curve with 10-ks resolution,
and investigated the Fe line flux. While the dynamic range of the
three flux slices are only a factor of 2, the line flux remains constant
between the flux slices, confirming the result obtained above. 
The iron K line is resolved and the inferred FWHM ($\sim 10000$ km
s$^{-1}$) implies a smaller line-emitting radius than the BLR (FWHI
5000 km s$^{-1}$, Ho et al 1997). The lack of significant variability
in the line flux is unexpected in this compact system. Additional studies of a wider range of ionizing flux variations will be
needed to verify the result.

%%%

\begin{table}
\begin{center}
\caption{Iron K line variability.}
\begin{tabular}{lccccc}
Data & Range & Exposure & Counts & $F_{\rm 3-10}$ & $I_{\rm FeK}$ \\
& (1) & (2) & (3) & (4) & (5) \\[5pt]
\multicolumn{6}{l}{TIME ORDER} \\
{\it i-1} & 0-90 & 40.0 & 5243 & 4.0 & $7.5^{+2.3}_{-1.9}$ \\
{\it i-2} & 90-160 & 29.8 & 2450 & 2.5 & $8.1^{+1.9}_{-2.2}$ \\
{\it i-3} & 160-230 & 29.0 & 4117 & 4.3 & $7.1^{+3.1}_{-2.1}$ \\ [5pt]
\multicolumn{6}{l}{FLUX ORDER} \\
L & $\leq 0.30$ & 51.8 & 3592 & 2.2 & $5.7^{+1.7}_{-1.5}$ \\
M & 0.30-0.55 & 31.7 & 4772 & 4.4 & $7.0^{+1.8}_{-2.6}$ \\
H & $\geq 0.55$ & 16.2 & 3681 & 6.7 & $9.4^{+3.4}_{-3.9}$ \\
\end{tabular}
\begin{list}{}{}
\item[(1)] in unit of $10^3$ s in the light curve
  in Fig. 1 for the time intervals, or \cps for the source count rate
  range
\item[(2)] $10^3$ s
\item[(3)] background-corrected integrated counts
\item[(4)] 3-10 keV flux in $10^{-12}$\ergpspsqcm
\item[(5)] Fe K line flux in $10^{-6}$\phpspsqcm
\item[Note:] The error in the line flux is of the order of $1\sigma $.
\end{list}
\end{center}
\end{table}

\section{Discussion}

\subsection{Origin of spectral variability}

Strong spectral variability is observed in the XIS band. The general
trend is that the spectrum is steeper when the source is brighter, and
vice versa, although a close inspection shows that it is not a simple
function of the source brightness (Fig. 11 and Table 3). The spectral
analysis of the six time intervals (Table 3) appears to show that the
spectral variability can be largely accounted for by continuum slope
changes. Variability in the warm absorber is less likely, but a good
alternative source of the variability and may also change the
spectrum. However, for the rapid changes observed in NGC4395, it is
generally difficult to draw strong conclusions using spectra of
the CCD spectrometer's resolution as to whether the warm absorber is
actually responsible for the variability.

The continuum slope can vary rapidly, as it is probably controlled by
the heating/cooling in the hot corona at small radii of the accretion
disk. Any change in the physical conditions of the corona can
propagate approximately in a light crossing time, which is a few tens of
seconds for the radius of 10 gravitational radius in NGC 4395. On the
other hand, the warm absorber would need a longer time, the order of
the recombination timescale, which depends on the density of the
absorbing matter, to respond to the variability of the illuminating
source. The X-ray colour HR2, derived from the 1-2.5 keV and 2.5-5 keV
data, would be sensitive to any changes in the warm absorber. Possible
delays in the response of the hardness ratio to the source flux
variation seen in some occasions may be a signature of variations in
the warm absrober. 

we now consider that this possible delay of $\simeq 2\times 10^4$ s is the response
time of the warm absorber. For an ionizing luminosity of $3\times
10^{40}$ \ergps, the calculation of the recombination
timescales of the relevant photoionized gas, using the formulae given in Otani et al
(1996) and Krolik \& Kriss (2001), yields the density of $n\sim 10^7$
cm$^{-3}$ and the distance of the absorbing matter $r\sim 3\times
10^{-3}$ pc. This density is relatively high but lower than that
inferred for the variable warm absorber in
NGC 4051 (Krongold et al 2007) in which the X-ray spectral variability
is attributed to variable absorption (but see Ponti et al
2006).

\subsection{Hard X-ray spectrum and broad-band energy distribution}

The hard X-ray spectrum of NGC4395 above 10 keV, measured with the
HXD-PIN, is $\Gamma\sim 2$, the typical value for Seyfert galaxies,
while the spectrum below 10 keV is much harder; $\Gamma\simeq 1.4$,
even after correcting for the absorption (Table 1). Although the PIN
slope is not well constrained, this might be indicative of a spectral break
around 10-20 keV or an extra absorption component that we failed to model.
A comparison with two INTEGRAL data points in the 20-40 keV and 40-100
keV bands (Fig. 13) suggests that the hard X-ray spectrum may actually
continue towards 100 keV with a slope $\Gamma\sim 1.5$, similar to
that obtained from the XIS data. However, since the INTEGRAL data were
not taken contemporaneously, this is still uncertain due to possible
flux variability. A stronger constraint on the hard X-ray spectral shape
would therefore be desirable. NGC 4395 is not in the Fermi LAT Bright Source
List \footnote{http://fermi.gsfc.nasa.gov/ssc/data/access/lat/bright\_src\_list}, which at least rules out such a hard spectrum
not extending to 100 MeV.

The multiwavelength spectral energy distribution (SED) is constructed
by taking photometric data collected from NED (Fig. 13; see also Moran
et al 1999; Lira et al 1999; Iwasawa et al 2000). Only the data taken
from a small aperture were used to minimize the contamination
from the host galaxy, which can be large for this low luminosity
nucleus. The IRAS data points are discarded since most of the IRAS
fluxes are believed to originate in the galaxy (Lira et al 1999).

Moran et al (1999) pointed out the anomalous SED shape of NGC4395,
which differs from that for the luminous QSOs (e.g., Elvis et al 1994) and also
from the low luminosity AGN (Ho 1999), as suggested by the absence of an
optical-UV bump and the radio quiet nature of the galaxy (the radio luminosity 5
orders of magnitude below the optical; see also Wrobel \& Ho 2006).
Dust reddening, which might suppress the optical light, was discussed
by Moran et al (1999), who did not find compelling evidence of
significant reddening. A remarkable feature of the SED in Fig. 13 is
that the peak of the energy output occurs at the hard X-ray range,
unlike normal Seyfert galaxies and QSOs, in which optical-UV emission
dominates the bolometric output. Without soft seed photons from the
accretion disk, it would be difficult to explain the production of
this strong hard X-ray emission with the thermal Comptonization model
(e.g., Haardt \& Maraschi 1991), unless the disk thermal emission
escapes our detection into the unobservable shorter wavelength UV
range, given the small black hole mass, e.g., the blackbody emission
peak would move by a half decade in wavelength, relative to Seyfert
galaxies with a black hole mass of $\sim 10^7 M_{\odot}$.

The distinctive SED shape of NGC4395 implies that the use of the
empirical relation to estimate the bolometric luminosity from the 5100
\AA\ luminosity, which is calibrated by the data for luminous AGN, may
be inappropriate. The estimated 0.5-100 keV luminosity is $\sim
4\times 10^{40}$ \ergps, which happens to be comparable to the
empirical estimate of the bolometric luminosity. If the bulk of the
bolometric luminosity indeed comes from the hard X-ray emission, the
Eddington ratio remains low, $10^{-3}$-$10^{-2}$, depending on the
$M_{\rm BH}$ estimate, and lies in an intermediate range between the
low accretion-rate (the Eddington ratio of $< 10^{-3}$) objects such as
M87 and normal Seyfert/QSOs. The radio quietness of the nucleus
suggests that the accretion flow is probably not in the form of
an advection-dominated type as proposed for objects with a much lower
accretion rate. However, we note that NGC4395 is located on the
``fundamental plane of black hole activity'' proposed by Merloni,
Heinz \& Di Matteo (2003). The fundamental plane is parameterised by
black hole mass, radio luminosity, and X-ray luminosity. Fitting this
relationship suggests that the dim optical emission could be an unusual
feature for this active nucleus.

\section*{Acknowlegments}
We thank all the members of the Suzaku team. We also thank Simon
Vaughan, Elisa Costantini and Pierre-Olivier Petrucci for useful
discussion and the referee for helpful comments. The data reduction and analysis were performed using the
software packages HEASoft provided by NASA's HEASARC. This research
made use of the NASA/IPAC Extragalactic Database (NED).

\end{document}